\documentclass[%
 reprint,
superscriptaddress,
 amsmath,amssymb,
 aps,
]{revtex4-2}

\usepackage{amsmath}
\usepackage{comment}
\usepackage{multirow}
\usepackage{array}
\usepackage{graphicx}% Include figure files
\usepackage{dcolumn}% Align table columns on decimal point
\usepackage{bm}% bold math
\usepackage{hyperref}% add hypertext capabilities
\usepackage{orcidlink}
\usepackage{cleveref}

\newcommand{\UCB}{\affiliation{Department of Physics, University of California, Berkeley, Berkeley, CA 94720, USA}}
\newcommand{\UNH}{\affiliation{Department of Physics \& Astronomy, University of New Hampshire, 9 Library Way, Durham NH 03824, USA}}
\newcommand{\WSU}{\affiliation{Department of Physics \& Astronomy, Washington State University, Pullman, Washington 99164, USA}}
\newcommand{\Caltech}{\affiliation{TAPIR, Walter Burke Institute for Theoretical Physics, MC 350-17, California Institute of Technology, Pasadena, California 91125, USA}}
\newcommand{\Cornell}{\affiliation{Cornell Center for Astrophysics and Planetary Science, Cornell University, Ithaca, New York, 14853, USA}}
\newcommand{\AEI}{\affiliation{Max Planck Institute for Gravitational Physics (Albert Einstein Institute), Am M{\"u}hlenberg~1, D-14476 Potsdam, Germany}}
\newcommand{\UIUC}{\affiliation{Department of Physics, University of Illinois at Urbana-Champaign, Urbana, Illinois, 61801, USA}}

\graphicspath{{./}{figures/}}

\begin{document}

\preprint{APS/123-QED}

\title{Stability of hypermassive neutron stars with realistic rotation and entropy profiles}% Force line breaks with \\

\author{Nishad Muhammed}\WSU
\author{Matthew D. Duez}\WSU
\author{Pavan Chawhan}\WSU
\author{Noora Ghadiri}\UIUC
\author{Luisa T. Buchman}\WSU
\author{Francois Foucart}\UNH
\author{Patrick Chi-Kit \surname{Cheong}}\UNH\UCB
\author{Lawrence E. Kidder}\Cornell
\author{Harald P. Pfeiffer}\AEI
\author{Mark A. Scheel}\Caltech

\date{\today}% It is always \today, today,

\begin{abstract}
Binary neutron star mergers produce massive, hot, rapidly differentially rotating neutron star remnants; electromagnetic and gravitational wave signals associated with the subsequent evolution depend on the stability of these remnants. Stability of relativistic stars has previously been studied for uniform rotation and for a class of differential rotation with monotonic angular velocity profiles. Stability of those equilibria to axisymmetric perturbations was found to respect a turning point criterion:  along a constant angular momentum sequence, the onset of unstable stars is found at maximum density less than but close to the density of maximum mass.  In this paper, we test this turning point criterion for non-monotonic angular velocity profiles and non-isentropic entropy profiles, both chosen to more realistically model post-merger equilibria. Stability is assessed by evolving perturbed equilibria in 2D using the Spectral Einstein Code. We present tests of the code's new capability for axisymmetric metric evolution. We confirm the turning point theorem and determine the region of our rotation law parameter space that provides highest maximum mass for a given angular momentum.

%\begin{description}
%You may use the \texttt{description} environment to structure %your abstract;
%use the optional argument of the \verb+\item+ command to give %the category of each item. 
%\end{description}
\end{abstract}

\maketitle

%\tableofcontents

\section{\label{sec:level1}Introduction \protect\\}

Binary neutron star mergers are important multimessenger astrophysical sources and probes of high density matter. Gravitational waves from the late-inspirals of such events have now been detected~\cite{LIGOScientific:2017vwq,LIGOScientific:2020aai}, in one case accompanied by electromagnetic counterparts~\cite{LIGOScientific:2017ync}. The high-frequency post-merger gravitational waveform and the electromagnetic signals (e.g. kilonova, gamma ray burst) are sensitive to the fate of the post-merger remnant.  This will be a hot, rapidly and differentially rotating star, which, depending on the binary mass and the equation of state, might collapse promptly to a black hole, might persist until secular evolution drives it to an unstable state followed by collapse, or might persist for longer times as a supramassive neutron star or indefinitely as a regular neutron star.  In delayed and no collapse cases, the remnant persists for many dynamical timescales, therefore in quasi-equilibrium configurations.  The presence and timescale of prompt or delayed collapse depends crucially on the stability of these equilibria to collapse.  (For review of binary neutron stars, see~\cite{Faber:2012rw,Baiotti:2016qnr,Burns:2019byj}.)

Although the stability of stellar equilibria is a classic problem~\cite{thompson1979stability, katz1978number,katz1979number, Sorkin:1981jc,Sorkin:1982,Friedman:1988}, the stability of hypermassive neutron stars is addressed in relatively few studies (e.g~\cite{Weih:2017mcw,Bozzola:2017qbu,Zhou:2019hyy,Espino:2019xcl}), and much remains unknown. Stability of relativistic stellar equilibria can be determined by finding the eigenfrequencies of linear perturbations or by full nonlinear numerical evolutions.  A way of evaluating stability from equilibria alone, without any sort of evolution, would be extremely helpful.  This explains interest in turning point methods, which provide information about stability from sequences of equilibria.  A sequence here means a one-dimensional slice in the space of equilibria, usually parameterized by the maximum baryonic density $\rho_{\rm max}$.  For arbitrary rotation, entropy, and composition profiles, this space would be infinite dimensional.

The turning point theorem~\cite{Sorkin:1982,Friedman:1988,Takami:2011zc} applies to uniformly rotating stars.  It assumes a one-parameter equation of state and, furthermore, that the pulsations of the star are governed by the same one-parameter equation of state.  Because uniform rotation is presumed to persist, it is a criterion for secular stability, i.e. stability on timescales on which uniform rotation is enforced.  The theorem applies to axisymmetric modes, the ones related to collapse, and does not address non-axisymmetric rotational instabilities.  [Indeed, many quasi-toroidal differentially rotating neutron stars are found to be unstable to non-axisymmetric (one-arm and bar mode) instabilities~\cite{Espino:2019xcl}.]  The space of equilibria is then two-dimensional, with total baryonic mass $M_0$ (the number of nucleons multiplied by a fiducial mass per baryon) and total angular momentum $J$ uniquely determining a star.  A constant-$J$ subspace is a 1D sequence.  If the total gravitational mass $M$ on the sequence has a maximum, then stars on the sequence at higher $\rho_{\rm max}$ are unstable.  The neutral point on the sequence separating secularly stable from unstable stars is at slightly lower  $\rho_{\rm max}$ for nonzero $J$~\cite{Takami:2011zc}.  Numerical evolutions find the dynamical stability neutral point to be close to the turning point~\cite{Shibata:1999yx,Weih:2017mcw}.

The turning point theorem does not apply to differentially rotating or non-isentropic stars, but Kaplan~{\it et al}~\cite{Kaplan:2013wra} conjecture that the turning point criterion remains approximately valid.  Their argument presumes that equilibrium $M$ depends to first order only on conserved quantities $M_0$, $J$, and total entropy $S$, and not on the angular momentum and entropy distributions.  They also note that only ``approximate turning points'' (not all conserved quantities having extrema at the same point on the sequence) are found in general, but they propose that this will be sufficient.

The stability of hypermassive neutron stars was studied, and Kaplan~{\it et al}'s conjecture tested, by Weih~{\it et al}~\cite{Weih:2017mcw} using numerical evolutions of these equilibria.  To construct equilibria, one must choose a rotation profile, and Weih~{\it et al} chose the j-constant law:
 \begin{equation}
 \label{eq:j-const}
     j(\Omega)=A^{2}\left(\Omega_{\mathrm{c}}-\Omega\right),
 \end{equation}
where $j$ is the specific angular momentum, $\Omega$ is the angular velocity, $\Omega_c$ is the central angular velocity, and $A$ is a free parameter with dimensions of length which controls the degree of the differential rotation. The name j-constant is chosen because in the Newtonian limit the specific angular momentum is constant~\cite{Galeazzi:2011nn,eriguchi:1985}.  

\begin{figure}
    \centering
    \includegraphics[width=1\linewidth]{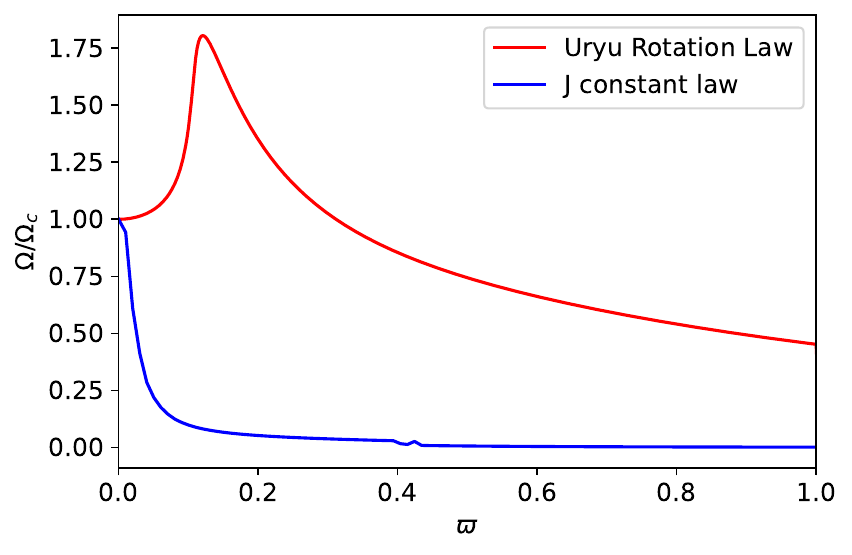}
     \label{fig:sample-profiles}
    \caption{Angular velocity $\Omega$ normalised to $\Omega_{c}$ as a function of coordinate distance from the axis $\varpi$ along the equator for two rotation laws. The top curve uses Eq.~\ref{eq:uryu-general} with $p=1$, $q=3$, $A=0.96$, $B=0.73$.  The bottom curve uses Eq.~\ref{eq:j-const} with $A=0.2$.}  
\end{figure}

Rotation profiles constructed with this law have $\Omega$ that monotonically decreases with distance from the center of the star. As Weih~{\it et al} themselves note, this is not a good match for the rotation profiles observed in remnants produced by binary neutron star merger simulations, which predict a non-monotonic $\Omega$ that peaks some distance away from the rotation axis~\cite{Kastaun:2014fna,Hanauske:2016gia,de2020numerical}.  Rotation laws that do capture this $\Omega(r)$ profile shape have been constructed by Ury{\={u}}~{\it et al}~\cite{Uryu:2017obi}.  The key idea is to specify $\Omega$ as a function of $j$ rather than vice versa.  In particular, one such profile is
\begin{equation}
\label{eq:uryu-general}
    \Omega\left(j ; \Omega_c\right)=\Omega_c \frac{1+\left[j /\left(B^2 \Omega_c\right)\right]^p}{1+\left[j /\left(A^2 \Omega_c\right)\right]^{q+p}},
\end{equation}
where $A$, $B$, $q$, and $p$ are specified constants.  An example of rotation profiles produced with the two laws is shown in Fig.~\ref{fig:sample-profiles}

In this paper, which may be considered an extension of Weih~{\it et al}'s study, we investigate the stability of hypermassive stars with non-monotonic angular velocity profiles.  Furthermore, we consider a range of (convectively stable) entropy profiles within the range plausible for binary neutron star mergers. We introduce a new 2D axisymmetric implementation of the Spectral Einstein Code for our numerical evolutions.  Our results vindicate the approximate turning point criterion.  In addition, we survey the parameter space of Ury{\={u}}~{\it et al} type rotation laws, seeing which values of the parameters are conducive to high maximum mass.

The organization of the paper is as follows. In Section~\ref{sec:methods}, we discuss the methods of building our initial data and carrying out evolutions. Next, in Section~\ref{sec:Sequences and Stability Testing}, we discuss the numerical experiments undertaken for this study. Results are presented and analyzed in Section~\ref{sec:results}. We summarize and conclude in Section~\ref{sec:summary}. We use the geometrized units, in which $c = G = M_{\odot} = 1$, unless stated otherwise.

\section{Equilibria and Evolution Methods}
\label{sec:methods}

\subsection{Equation of state and entropy profile}

The matter in the star is modeled as a perfect fluid with stress-energy tensor
\begin{equation}
  T^{\mu\nu} = \rho h u^{\mu}u^{\nu} + Pg^{\mu\nu}\ ,
\end{equation}
where $\rho$ is the baryonic density, $h$ is the
specific enthalpy, $u^{\mu}$ is the 4-velocity, and
$P$ is the pressure. The neutron star matter is modeled using the DD2 equation of state~\cite{DD2}.  DD2 provides $P$ and $h$ as functions of baryonic density $\rho$, temperature $T$, and reduced electron fraction $Y_e$.  It is based on a relativistic mean field model and is publicly available in tabulated form at \url{http://www.stellarcollapse.org}~\cite{OConnor2010}.  It predicts radius $R_{\rm NS}= $13.1\,km and tidal deformability $\Lambda=860$ for a 1.35\,$M_{\odot}$ neutron star.

Our algorithm for constructing equilibrium models requires one-dimensional equations of state (EOS):  $P=P_{\rm eq}(\rho)$. The 1D EOS we use for equilibrium construction are one-dimensional cuts of DD2, created by imposing two conditions to determine $Y_e$ and $T$ for each $\rho$. The first condition is beta equilibrium:  $\mu_p+\mu_e=\mu_n+\mu_{\nu}$, where we take the electron neutrino chemical potential $\mu_{\nu}$ to be zero. The second condition is an explicit choice for the dependence of specific entropy $s$ on density: $s=s_{\rm eq}(\rho)$.  We also produce one EOS, \texttt{ColdStar}, for which the temperature is $T = 0.01\text{MeV}$, the table minimum temperature. Based on the choice of $s_{\rm eq}(\rho)$, we have the following nomenclature for the EOSs: \texttt{CE1} corresponds to constant specific entropy, $s=1\ \text{k}_{B}\text{/baryon}$, \texttt{CE2} corresponds to $s=2.2\ \text{k}_{B}\text{/baryon}$. \texttt{VE1} is a variable entropy cut motivated by the thermodynamic profile of the merger remnant in Perego~{\it et al.}~\cite{Perego_etal}. It has specific entropy varying between $10^{-3}$-- 6 $\text{k}_{B}\text{/baryon}$ for NS density range $10^{12}$-$10^{16}$ $\text{gm}/\text{cm}^3$. \texttt{VE2} has entropy varying between 3 $\text{k}_{B}\text{/baryon}$ and 1 $\text{k}_{B}\text{/baryon}$ for the same density range.

\begin{figure}
    \centering
    \includegraphics[width=1\linewidth]{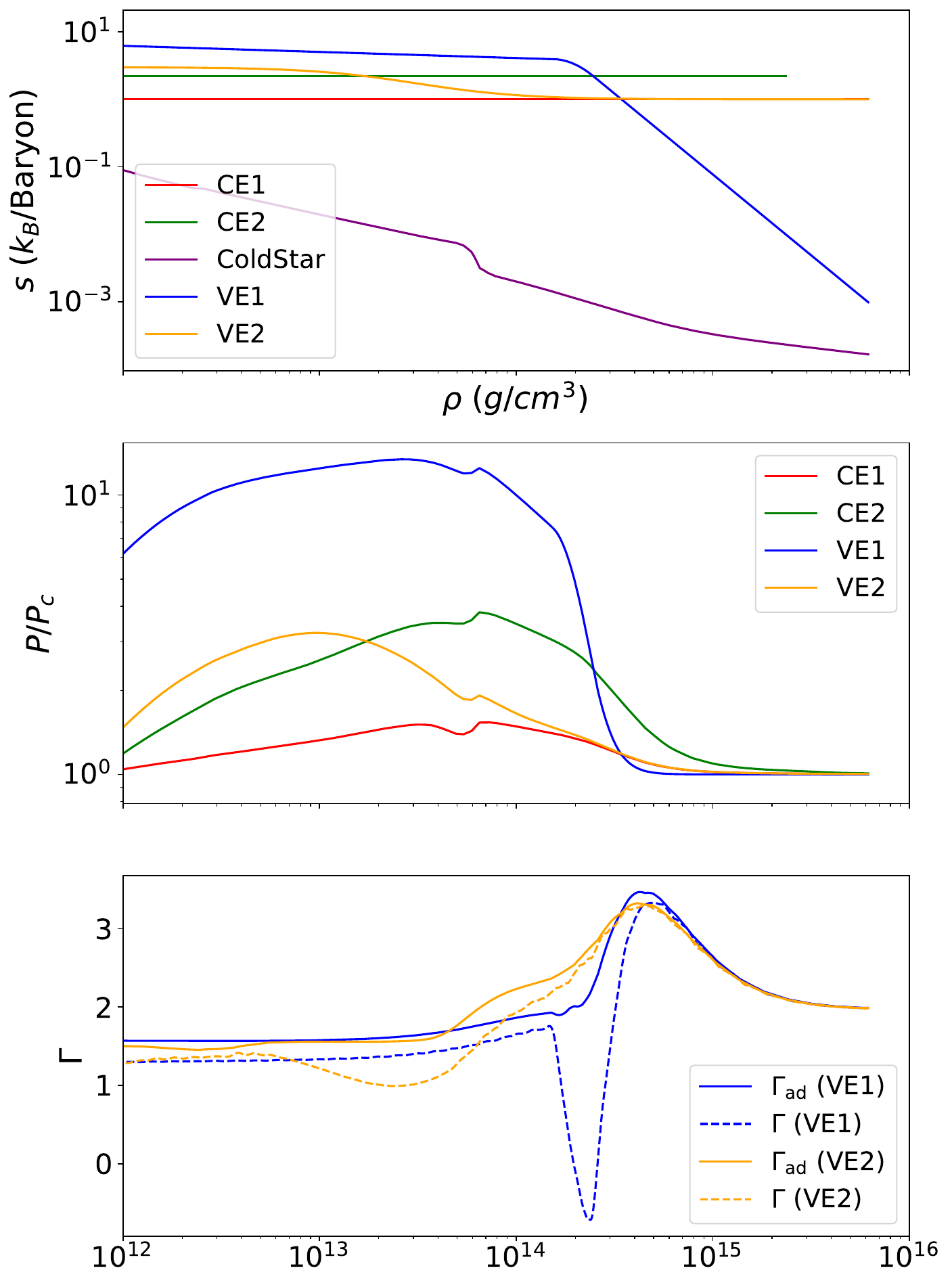}
    \caption{Profiles of different EOS cuts plotted for density range $10^{12}$-$10^{16}$ $\text{gm}/\text{cm}^3$. \textbf{Top panel:} $\Gamma$ against density. \textbf{Center Panel:} $P/P_c$ against density, where $P_c$ is the pressure corresponding to \texttt{ColdStar}. \textbf{Bottom Panel:} Specific entropy against density.}
    \label{fig:1DEOS}
\end{figure}

Profiles of EOS cuts are shown in Fig.~\ref{fig:1DEOS}, where we plot three decades of density up to the highest neutron star maximum density, the range relevant to the structure of our stars. Comparison of $P$ to $P(T=0.01\rm MeV)$ indicates the degree of degeneracy; we see that the cores are always degenerate but the envelopes are not (reflecting the expected outcome of mergers). Because of entropy and composition ($Y_e$) gradients in the equilibrium star, when perturbed, these stars will move to regions of the equation of state space outside the cut used to construct equilibria, another way in which our stars fall outside the domain of turning point theorems. Therefore, we also compare $\Gamma\equiv d\ln P/d\ln\rho|_{s=s_{\rm eq}(\rho)}$ to $\Gamma_{\rm ad}\equiv d\ln P/d\ln\rho|_{s=\rm const}$ to indicate the strength of buoyancy forces in the non-isentropic cases.

The entropy profile for \texttt{VE1} has a sharp change in slope, which leads to a density regime of very shallow $P$ vs. $\rho$.  In fact, $dP/d\rho$ is actually slightly negative in the range $2-2.5\times 10^{14}\,\text{gm/cm}^3$, which can be seen in the plot of $\Gamma$.  (Note that this is neither an isothermal nor an adiabatic derivative; the fluid is thermodynamically stable, and sound waves are stable.)  In practice, the equilibrium solve ``jumps over'' this density region, so $P(\rho)$ is effectively flat, reminiscent of a first-order phase transition, and the resulting stellar profiles have an abrupt jump in density.  Although inadvertent, this feature allows us to test the turning point criterion for equilibria with density jumps, a feature which might appear in post-merger remnants if a first-order phase transition from hadronic to quark matter is present~\cite{Shuryak:1980,Bauswein:2018bma,Most:2018eaw,Weih:2019xvw,Prakash:2021wpz}.

%For \texttt{VE1}, we see a negative value of $\Gamma$ around $\rho = 2 \times 10^{14} \text{gm/cm}^3$, implying negative $dp/d\rho_0|_{s=s_{\rm eq}(\rho)}$. The pressure starts decreasing  around $\rho = 2 \times 10^{14} \text{gm/cm}^3$, and again starts increasing around $\rho = 2.5 \times 10^{14} \text{gm/cm}^3$. For a system in hydrostatic equilibrium, densities close to and greater than $\rho = 2 \times 10^{14} \text{gm/cm}^3$ would never be attained as one goes from low density to high density. Instead, there would be an abrupt jump in density resembling a first-order phase transition.

\subsection{Rotation profile and construction of equilibria}

We produce axisymmetric equilibrium configurations using the
code of Cook, Shapiro, and Teukolsky~\cite{Cook:1992,Cook:1994}, which we call ``RotNS''.

The spacetime metric is written in the form
\begin{eqnarray}
  ds^2 &=& -e^{\gamma+\lambda}dt^2 + e^{2\alpha}(dr^2 + r^2 d\theta^2) \\
  \nonumber
  & & + e^{\gamma-\lambda}r^2\sin^2\theta (d\phi^2 - \omega dt)^2.
\end{eqnarray}

The fluid motion is taken to be azimuthal, so the proper velocity $v$, the Lorentz factor $u^t$, and the specific angular momentum $j$ are
\begin{eqnarray}
  v &=& (\Omega - \omega)r\sin\theta e^{-\lambda}, \\
  u^t &=& (1-v^2)^{-1/2}e^{-(\rho+\gamma)/2}, \\ 
  j &\equiv& u^tu_{\phi} = \left(u^t\right)^2e^{\gamma-\lambda}
  r^2\sin^2\theta (\Omega-\omega).
\end{eqnarray}

An integrability condition on the equation of hydrostatic equilibrium requires we choose for the rotation law either uniform rotation ($\Omega=$ constant) or that $j=j(\Omega)$ or $\Omega=\Omega(j)$.  The original RotNS used the law $j(\Omega)=A^2(\Omega_c-\Omega)$ for constant $A$.  This law does not allow non-monotonic rotation profiles of the sort seen in binary neutron star simulations.  Such profiles can be constructed if $j$ is taken to be the independent variable.  Thus, following Ury{\={u}}~{\it et al}~\cite{Uryu:2019}, we implement the following rotation law,
\begin{equation}
\label{eq:uryu}
  \Omega(j;\Omega_c) = \Omega_c\frac{1+j/(B^2\Omega_c)}{1+[j/(A^2\Omega_c)]^4}, 
\end{equation}
i.e. we choose $p=1$, $q=3$ from the more general law, Eq.~\ref{eq:uryu-general}. A typical profile is shown in Fig.~\ref{fig:sample-profiles}.

Given $\Omega_c$, one can find $j$ at any point $(r,\theta)$ by finding the root of $f_j(j)\equiv u^t(\Omega[j])u_{\phi}(\Omega[j])-j=0$. Given $j$ at each point, the matter distribution is given by
the Bernoulli integral
\begin{eqnarray}
  H(\rho) &=& H_{\rm eq}(j;\Omega_c) \\
  \nonumber
  &=& \frac{1-\zeta}{\sqrt{1-v^2}}
  \exp[-(\lambda+\gamma)/2 + I(j;\Omega_c)]\ ,
\end{eqnarray}
where $H$ is the non-isentropic generalization of the specific enthalpy~\cite{Camelio:2019rsz}
\begin{equation}
  \ln(H) \equiv \int_0^P \frac{dP}{\rho h}\ .
\end{equation}
Here the integral is taken along the curve $s=s_{\rm eq}(\rho)$, $\zeta$ is an integration constant, and the rotation profile integral
\begin{equation}
  I(j;\Omega_c) = \int_0^j j'\frac{d}{dj'}\Omega(j';\Omega_c)dj'\ ,
\end{equation}
is messy but analytic.

Let us call the equatorial radius $r_e$, the polar radius $r_p$, the maximum density $\rho_{\rm max}$ and its coordinate distance from the axis $r_m$.  To find a model for a single equilibrium, RotNS specifies $\rho_{\rm max}$ and the ratio $\hat{r}_p\equiv r_p/r_e$.  In addition to solving for the metric, one needs to determine the appropriate constants $\Omega_c$ and $\zeta$.  This is done by an iterative process of refining an initial guess.  For the first $\rho_{\rm max}$ of each sequence, we start with a TOV star and then adjust $\hat{r}_p$ downward until the angular momentum $J$ reaches the desired value $J_{\rm seq}$, a 1D root find for $J(\hat{r}_p)-J_{\rm seq}$.  For the next $\rho_{\rm max}$, the star on the sequence for the previous $\rho_{\rm max}$ serves as the initial guess.

The procedure for determining the global constants is the following straightforward generalization of the original RotNS.  First, we define scaled metric potentials $\hat{\rho}\equiv\rho r_e^{-2}$,
$\hat{\gamma}\equiv\gamma r_e^{-2}$,
$\hat{\alpha}\equiv\alpha r_e^{-2}$.  These shall be taken as fixed for the relaxation procedure.
%Also fixed is the scaled location of density maximum $\hat{r}_m\equiv r_m/r_e$.

At the pole, $\Omega=\Omega_c$, $v=j=0$, $h=1$, so
\begin{eqnarray}
    1 &=& H_{\rm eq}(j=0,r=0; \Omega_c) \\
    \nonumber
      &=& (1-\zeta)e^{(\rho_p+\gamma_p)/2+I(0;\Omega_c)},
\end{eqnarray}
which provides an equation for $\zeta$. There are also two equations
at the equator ($r=r_e$, $\theta=\pi/2$, $j=j_e$) and another two at the point of maximum density ($r=r_m$, $\theta=\pi/2$, $j=j_m$).  These equations are 
\begin{eqnarray}
\label{eq:heq}
1 &=& H_{\rm eq}(j=j_e,r=r_e,\theta=\pi/2; \Omega_c), \\
\label{eq:jeq}
j_e &=& [u^tu_{\phi}](j=j_e,r=r_e,\theta=\pi/2; \Omega_c), \\
H(\rho_{\rm max}) &=& H_{\rm eq}(j=j_m,r=r_m,\theta=\pi/2; \Omega_c), \\
j_m &=& [u^tu_{\phi}](j=j_m,r=r_m,\theta=\pi/2; \Omega_c).
\end{eqnarray}
We solve these using Newton's method for the global parameters  ($\Omega_c$, $j_m$, $j_e$, $r_e$). When the maximum density is at the center, we use instead a 2D root finder, solving equations~\ref{eq:heq}~\ref{eq:jeq} for ($\Omega_c$,$j_e$).

The parameters $A$ and $B$ in Eq.~\ref{eq:uryu} must also be specified. In some sequences below, we take them to be constant. Alternatively, we can fix the ratios $\Omega_{\rm max}/\Omega_c$ and $\Omega_e/\Omega_c$, where $\Omega_{\rm max}$ is the maximum value of $\Omega$ (note: not the value of $\Omega$ where $\rho=\rho_{\rm max}$), and $\Omega_e$ is the equatorial $\Omega$.  Given these ratios, $A$ and $B$ can be determined by a 2D root find, which does not converge for desired angular momentum if the solve for $A$ and $B$ is performed within the relaxation for a single model. Instead, the solver for $A$ and $B$ must be the outer stage of the relaxation. If precise values of the ratios are not needed, a close approximation is obtained by solving for $A$ and $B$ at the completion of each successful new model, assuming one takes fairly small steps in $\rho_{max}$.
Fixing angular velocity ratios is what was used for the sequences in Ury{\={u}}~{\it et al}~\cite{Uryu:2019}.

We stress that there is no correct or (known) physically realistic choice in how to specify $A$ and $B$.  This is part of the ambiguity inherent in defining sequences and attempting to apply a turning point criterion in a parameter space of differentially rotating stars, something that is less likely to be noticed when using a rotation law family (like the j-constant family) with only one differential rotation parameter, which it might seem natural to hold constant.  Whether this ambiguity turns out to be important in predicting stability is something we address in this study.

\subsection{Numerical evolution}

\begin{figure*}
\includegraphics[width=1\columnwidth]{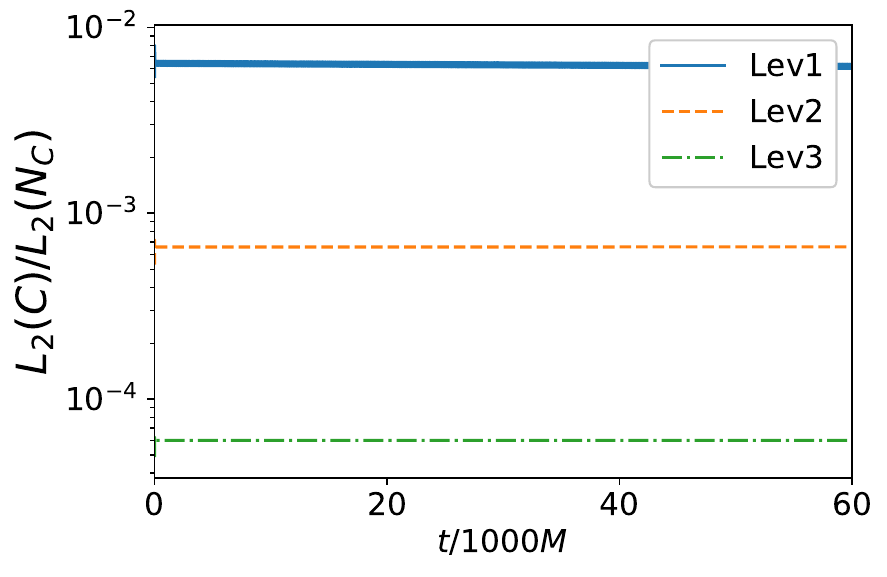}
\includegraphics[width=1\columnwidth]{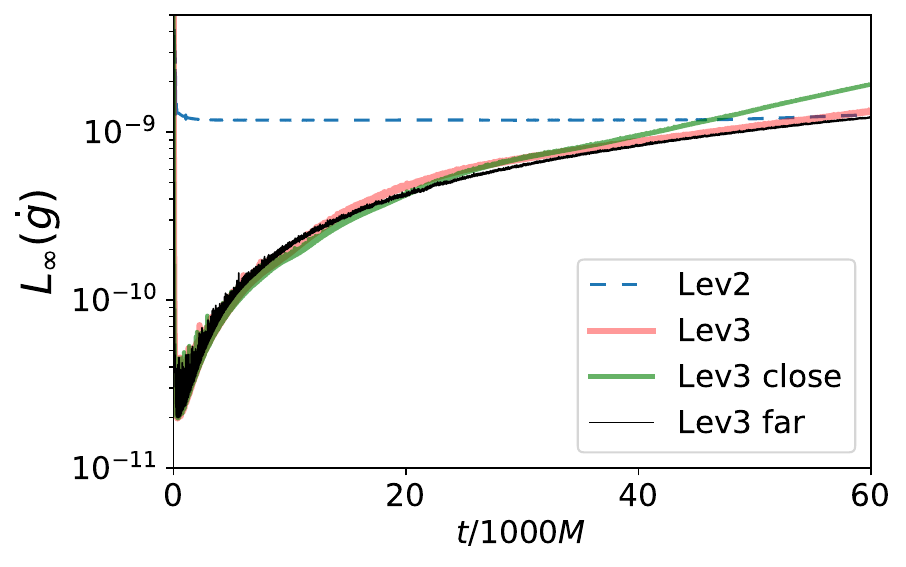} \\
\includegraphics[width=1\columnwidth]{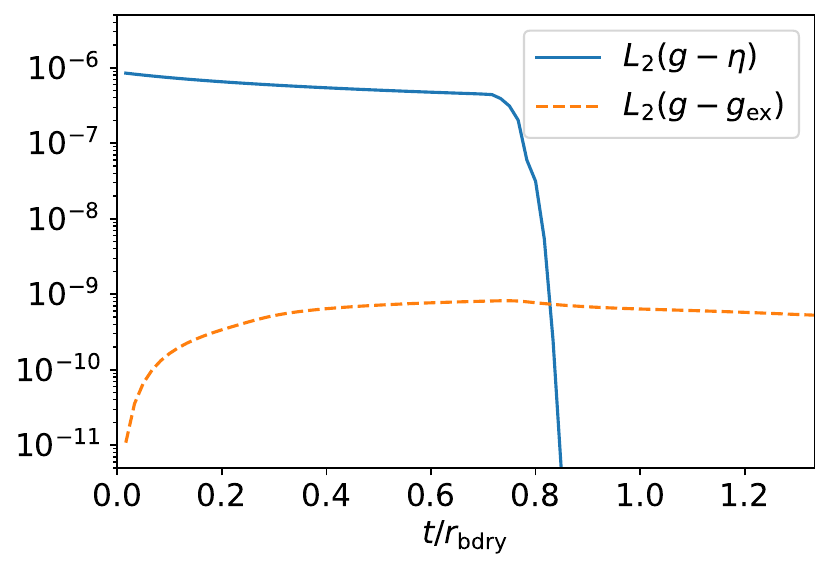}
\includegraphics[width=1\columnwidth]{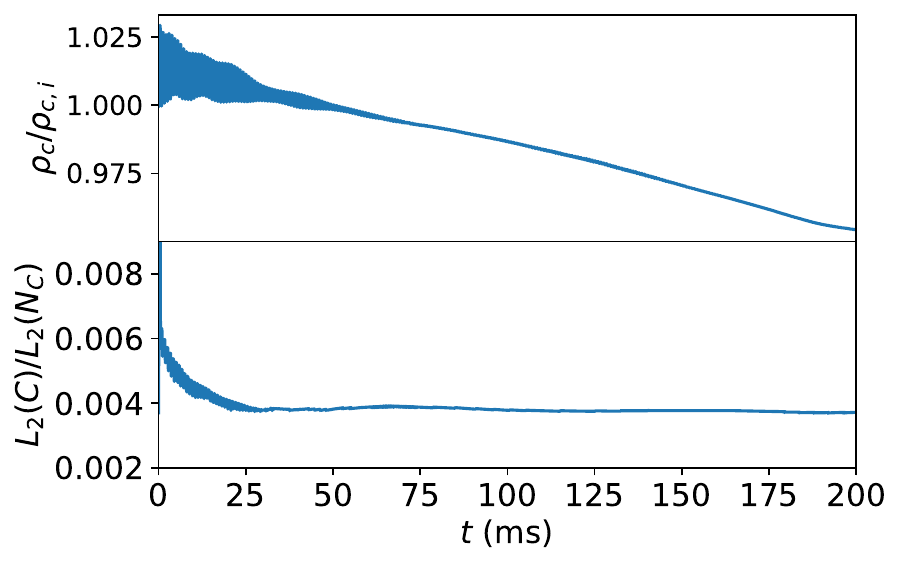}
\caption{Tests of SpEC's 2D metric evolution code.  {\bf Top left: } The constraint violation for the evolution of an isolated Kerr black hole at three resolutions.  {\bf Top right: } the norm on the time derivative of the metric.  (The lowest resolution is not shown, because its size of $\sim 10^{-5}$ would throw off the scale for the other resolutions, even on a logarithmic plot.)  For the higher resolution, we show three outer boundary locations, with the one with outer boundary closer than all the others (``Lev3 close'') clearly having worse late-time behavior.  Although it is not clearly visible on the logarithmic plot, the farthest boundary run (``Lev3 far'')has distinctly lower late-time slope than the next closest boundary (``Lev3''), and the lower resolution run (``Lev2'') has a linear growth like the Lev3 runs that become apparent at late times.  {\bf Bottom left: } Error for an axisymmetric gravitational wave.  The sharp drop in $L2(g-\eta)$ (the deviation of the metric from Minkowski) is when the wave reaches the outer boundary at $r=r_{\rm bdry}$ and passes through.  {\bf Bottom right: } an extended evolution of a differentially rotating star, with central density on top and constraint violation below.}
\label{fig:test-2d-evolutions}
\end{figure*}

We evolve using the Spectral Einstein Code (SpEC)~\cite{SpEC2024}. SpEC evolves the fluid using a conventional high-resolution shock capturing finite difference method, and it evolves the spacetime metric in the generalized harmonic formulation using a multidomain pseudospectral method.

Because the unstable mode triggering radial collapse is expected to be axisymmetric, and because we evolve many equilibria, we use the 2D axisymmetric version of SpEC.  Our multipatch 2D hydrodynamics code is described in detail in Jesse~{\it et al}~\cite{Jesse:2020}.  In that work, we had not developed an axisymmetric version of the pseudospectral metric evolution, so we were forced to evolve the metric on a 3D grid of colocation points for those applications with dynamical spacetimes.  Although inefficient, this method allowed accurate simulations of differentially rotating neutron stars for tens of milliseconds, but eventually such simulations succumbed to accumulating growth of violations in the constraint equations.

Here we introduce a new fully 2D version of SpEC, with metric evolution now carried out on a 2D grid representing a meridional cut through the presumed axisymmetic system.  Derivatives in the spatial direction perpendicular to the evolution plane are computed using the Cartoon method~\cite{Pretorius:2004jg,Hilditch:2015aba}.  The Cartoon method solves the symmetry condition for tensor $T$:  $\mathcal{L}_{\partial/\partial\phi}T=0$.  (As Hilditch~{\it et al} explain, this technique can be applied essentially unchanged even for spatial derivatives of the metric $\partial_kg_{\alpha\beta}$ even though this is not a covariant tensor.)

Given axisymmetry, only one side of the axis on a 2D cut needs to be evolved; the other is determined by the symmetry and can be replaced by appropriate symmetry/regularity boundary conditions on the axis. For this project, we have taken the algorithmically simpler path of evolving both sides.  The spectral grid is then constructed of concentric circular wedge domains (corresponding to Chebyshev radial basis functions and Fourier angular basis functions) with a filled shape in the center (corresponding to Matsushima-Marcus basis functions~\cite{Matsushima-Marcus:1995}).  We choose angular colocation points such that no points lie on the axis, and the grid is exactly symmetric across the axis.

In principle, roundoff error could lead to a breakdown of the symmetry across the axis.  This does not appear to be an issue in our simulations, but we have carried out simulations which enforce symmetry by replacing each component $g_{\mu\nu}$ and $\partial_{\alpha}g_{\mu\nu}$ after each time step with the average of the component on both sides of the axis (with appropriate symmetry factors).  This is inexpensive because symmetric pairs of points lie on the same domain and thus on the same processor.  We see little difference with or without averaging but have used it for the simulations reported here.

Fig.~\ref{fig:test-2d-evolutions} shows the results of some tests of the new code.  First, we perform a very long-time evolution of a Kerr-Schild black hole with spin $a/m=0.5$.  The system should be stationary but involves strong curvature.  On the top left, we plot the violation of the generalized harmonic constraints $C$, normalized to the size of the terms in the constraints $N_C$.  For each, we compute a volume integral $L_2$ norm:  $L_{2}(u)\equiv \sqrt{\left(\int u^2 dV\right)/\left(\int dV\right)}$  Constraint plots see rapid convergence with resolution and a quick settling to numerical equilibrium.  Metric components show an appearance of stasis.  To see continued evolution, we also plot the $L_{\infty}$ norm of $\sqrt{\Sigma_{\alpha\beta}\partial_tg_{\alpha\beta}^2}$, where the time derivative is computed from the difference in the metric between consecutive timesteps.  This converges rapidly with resolution at early times.  At very late times, it begins to grow linearly.  This might mean eventual problems for these simulations (if 
 the growth does not saturate before becoming visible), but the slope decreases with increasing distance to the outer boundary, so the growth of non-stationarity can be controlled and delayed in this way.

 For a dynamical vacuum problem, we evolve a radially outgoing $\ell=2$, $m=0$ gravitational wave packet.  The packet initial amplitude is a Gaussian with width 1.5 and peak amplitude $10^{-4.5}$ centered at distance $r=15$ from the origin.  The outer boundary for the run shown in the figure is at $r=60$.  The wave propagates to the boundary and leaves the grid, with $g_{\alpha\beta}$ remaining very close to the analytic solution of the linearized Einstein equations.  We demonstrate this by comparing $L_2(g_{\alpha\beta}-g^{\rm lin}_{\alpha\beta})$ (our error, plus small effect of nonlinearities) to $L_2(g^{\rm lin}_{\alpha\beta}-\eta_{\alpha\beta})$ (a measure of the strength of the wave).  Eventually, long after the wave has passed, constraint violations begin to grow and eventually spoil the simulation, but again, this can be delayed--apparently without limit--by moving the outer boundary sufficiently outward.

 Both of these applications indicate a need to improve the outer boundary condition in 2D simulations (although it is not clear why SpEC's constraint-preserving, frozen $\Psi_0$ boundary conditions do not work as well here as in 3D) for future studies involving very long evolution times.

 By contrast, our final test of a differentially rotating star evolves for a very long time with no sign of eventual trouble.  For this non-vacuum test, we use the model reported in Jesse~{\it et al}~\cite{Jesse:2020}, although here with no viscosity, so that the solution is expected to be stationary.  In Jesse~{\it et al}~\cite{Jesse:2020}, the fluid was evolved in 2D but the metric in 3D.  The star has an initial baryonic rest mass of 2.64 $M_{\odot}$ and an equatorial radius $R_e = 10.2$ km.  Its EOS is a 2-component piecewise polytrope.  The initial rotation profile for the star is given by $u^t u_\phi = \hat{A}(\Omega_0 - \Omega)$ where $\Omega_0$ is the angular velocity along the rotation axis, and we choose $\hat{A} = 0.8 R_e$.  Convergence of equilibria is demonstrated below in Section~\ref{sec:results}.  Here we check our ability to evolve stably for $>100$\,ms with dynamical metric evolution, something we could not achieve with 2D fluid grid and 3D metric grid.  We see stable constraints and a slow downward drift in central density, presumably because of numerical heating.  This 200\,ms evolution is much longer than we were able to evolve with 2D fluid, 3D metric evolution.  Evolutions with 2D fluid, 3D metric evolution usually develop unstable constraint violation growth, but there is no sign of this in the fully 2D stellar evolution.

\section{Sequences and Stability Testing}
\label{sec:Sequences and Stability Testing}

For this study, we have evolved more than 200 models on 23 different constant angular momenta sequences. The initial data for these models are characterized by different angular momenta, 4 different 1D EOS (encompassing non-isentropic variants) and two distinct rotation laws. The angular momenta, EOS details, parameters of Ury{\={u}} rotation law (A,B)  and names of these sequences have been documented in Table~\ref{table:SequenceNames}. The stability tests are conducted on a dynamical timescale spanning a few oscillation periods. This is adequate for assessing stability, as the collapse of a dynamically unstable hypermassive neutron star into a black hole occurs within this timescale. As a result, no viscous or radiation effects are included.

\begin{table}[]
\centering
\begin{tabular}{ |c|c|c|c|c| } 
\hline
\label{table:SequenceNames}
EOS & J & A & B & Name \\
\hline
\multirow{6}{2em}{CE1} & 6 & 0.79 & 0.55 & A\\ 
& 6.5 & 0.79 & 0.55 & B\\ 
& 8.5 & 1.45 & 1.48 & C\\ 
& 11 & 1.61 & 1.64 & D\\
& 10 & 1.22 & 0.92 & E\\
& 11 & 1.10 & 0.83 & F\\
\hline
\multirow{8}{2em}{CE2} & 11 & 0.96 & 0.73 & G\\
& 10 & 0.87 & 0.60 & H\\
& 10 & 0.91 & 0.63 & I\\
& 10 & 0.82 & 0.57 & J\\
& 11 & 0.89 & 0.62 & K\\
& 11 & 0.88 & 0.60 & L\\
& 9 & 0.87 & 0.60 & M\\
& 9 & 1.03 & 0.72 & N\\
\hline
\multirow{3}{2em}{VE1} & 10 & 0.65 & 0.45 & O\\
& 7 & 0.65 & 0.45 & P\\
& 4 & 0.65 & 0.45 & Q\\
\hline
\multirow{4}{2em}{VE2} & 10 & 1.08 & 0.75 & R\\
& 7 & 0.98 & 0.68 & S\\
& 7 & 1.08 & 0.75 & T\\
& 4 & 0.93 & 0.64 & U\\
\hline
\end{tabular}
\caption{List of constant-J sequences in this study. From left to right, the columns represent the entropy profiles, angular momenta, Ury{\=u} rotation law parameters A, B and sequence names respectively.}
\end{table}

The fluid grid is evolved on a 2D uniform Cartesian grid, covering a square that includes the star.  The metric is evolved on a separate 2D pseudospectral grid.  The pseudospectral grid includes a circle at the center surrounded by concentric annuli. The circle and the annuli all have the same angular resolution. The outer annuli are chosen to have larger radial extent than the inner ones to allow grid to be concentrated inside the star.

 We carried out tests on unperturbed models to determine adequate grid resolution. Six different resolutions for the fluid and pseudospectral grid were used, labeled ``Res 1'' through ''Res 6'', with Res 1 the lowest and Res 6 the highest resolution. It should be noted that these resolutions are distinct from the resolutions for 2D dynamical metric evolution discussed in Section~\ref{sec:methods}.

\begin{figure}
    \centering
    \includegraphics[width=1\linewidth]{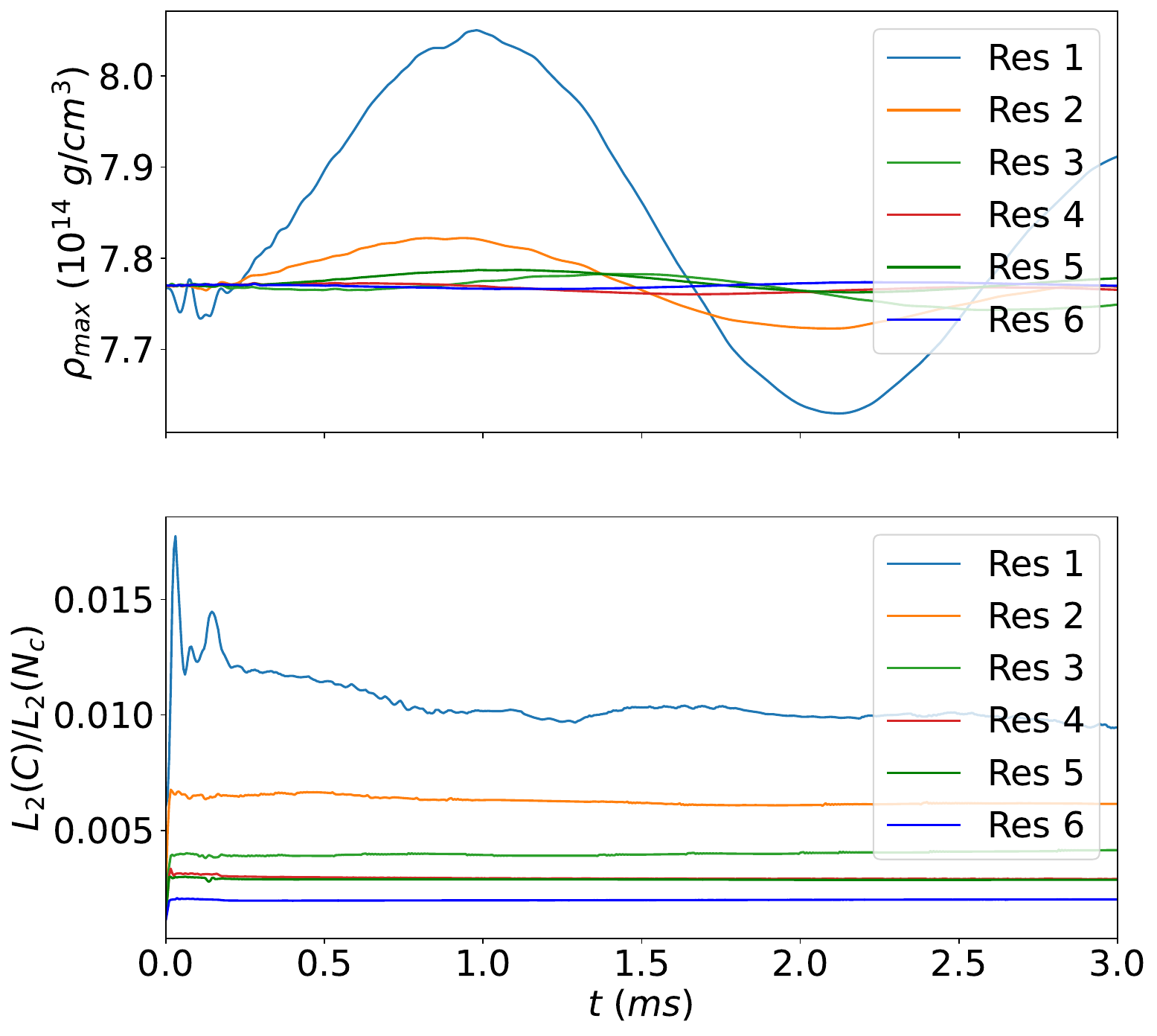}
    \caption{Convergence tests for a sample hypermassive star with a maximum energy density of $8.897\times10^{14}\ g/cm^3$ from sequence F (see Table~\ref{table:SequenceNames}). Res 1 through Res 5 represent low to high resolution respectively. The optimal grid resolution found was Res 3. {\bf Top panel:} maximum baryonic density vs time; {\bf Bottom panel:} normalized error of the generalized harmonic constraints.}
    \label{fig: GhCe_Resolution}
\end{figure}

 Fig.~\ref{fig: GhCe_Resolution} demonstrates the effect of resolution on evolution.  We plot maximum density and the $L_2$-norm of generalised harmonic constraint ($L_{2}(C)/L_{2}(N_{c})$) against time.  We see convergence toward stationarity and constraint satisfaction at low resolutions, but at sufficiently high resolution, deviation from equilibrium and constraint violation are dominated by the finite error of the RotNS initial data. As is evident from the figure, increasing the resolution beyond Res 3 does not have a significant impact on maximum density and $L_{2}(C)/L_{2}(N_{c})$. The resolution of the Res 3 fluid grid is $300^2$. Its pseudospectral grid has one circle followed by five concentric annuli with angular extents of 50 for all of them.  The total number of radial layers of colocation points (unevenly spaced, as mentioned above) is 564, extending out to a distance of 2940\,km.
 
Once the resolution was decided, tests with perturbation were executed. We introduce inward radial velocity perturbation:

\begin{equation}
\label{eq:perturbation}
\begin{aligned}
    \delta v_x(x, y, z)=\mathcal{A}|x|, \\
    \delta v_y(x, y, z)=\mathcal{A}|y|,
\end{aligned}
\end{equation}
where $\delta v_x$ and $\delta v_x$ is the velocity perturbation in the $x$ and $y$ direction respectively. $\mathcal{A}$ is a free parameter that sets the amplitude of the perturbation.  We tried different values for $\mathcal{A}$ and chose a value with density oscillations slightly smaller in amplitude than the spacing between adjacent evolved configurations on the sequence, so that the perturbation does not push the lowest-density unstable star to its adjacent stable configuration~\cite{Weih:2017mcw}.  The chosen amplitude was $\mathcal{A}$ = $-0.005$.  A perturbation of this amplitude does not increase the initial constraint violation.  Furthermore, we have checked that varying $\mathcal{A}$ does not alter which stars are stable and which unstable for a sample sequence.
 
\begin{figure}
    \centering
    \includegraphics[width=1\linewidth]{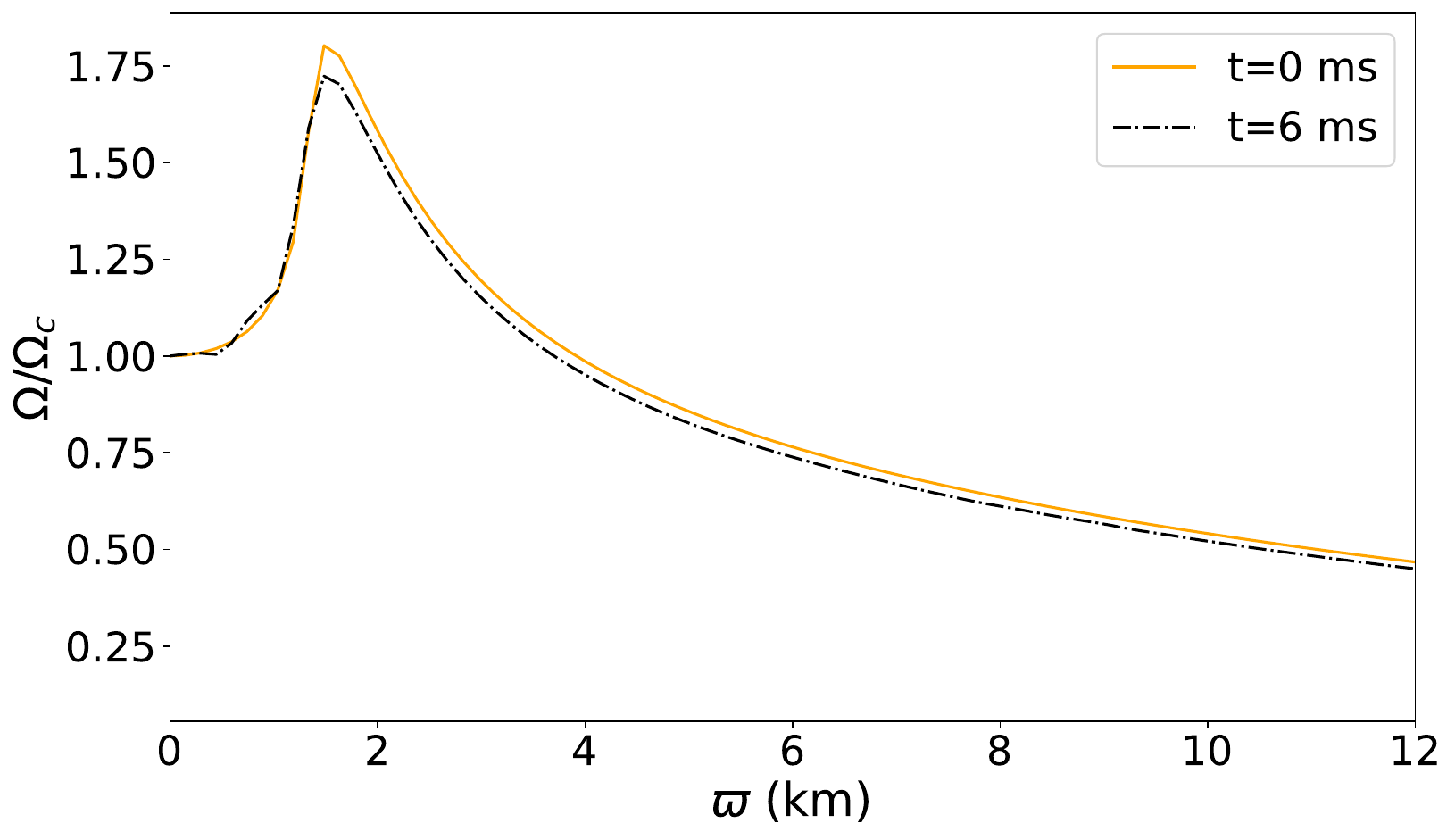}
    \caption{Angular velocity $\Omega$ as a function of coordinate dis-
tance from the rotation axis $\varpi$ along the equator at $t=0$ ms and $t=6$ ms with A = 0.96 and B = 0.73.}
    \label{fig: OmegaProfileEvolution}
\end{figure}

Fig.~\ref{fig: OmegaProfileEvolution} illustrates the rotation profile of sequence G at $t=0$\,ms and at $t=6$\,ms (at the end of the dynamical evolution). Note that the $\Omega$ does not monotonically decrease with radius but has a peak between the center and the equator. Since we are simulating post-merger remnant-like stars, we use $\Omega_{\rm max}/\Omega_{\rm c}$ in the range 1.5 - 2.1 and $\Omega_{\rm eq}/\Omega_{\rm c}$ in the range 0.3 - 0.8 for our evolutions~\cite{de2020numerical}.
The rotation profile is preserved throughout the evolution, indicating the model is in equilibrium.  This is in contrast to what was found for some toroidal stars with this rotational profile evolved under the assumption of conformal flatness~\cite{Cheong:2024rvk}.  The figure shows $\Omega$ of a model with $\rho_{max}$ = $7.3\times10^{15}$ g\ cm${}^{-3}$. We checked $\Omega$ for higher energy density stable stars on the sequence, and they show the same stationarity. Furthermore, the density profile also does not change much (less than 4\% at all points on the axis).

\section{Results}
\label{sec:results}

\subsection{Stability}
\label{stability}

\begin{figure}
    \centering
    \includegraphics[width=1\linewidth]{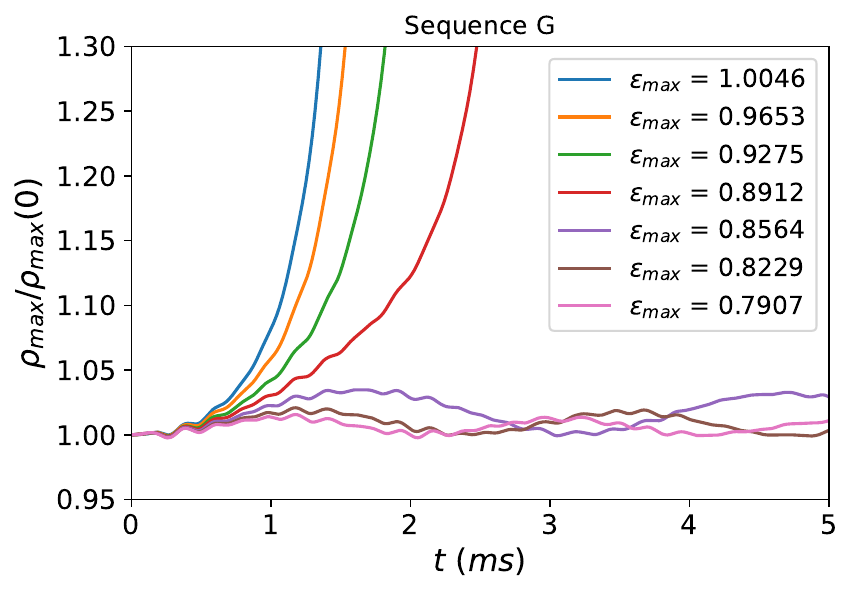}
    \caption{Evolution of the maximum baryonic density normalized to its initial value. All models are from the same sequence G with $A = 0.96$ and $B=0.73$. The labels indicate the maximum energy densities from sequence G in $10^{15}\ g/cm^3$ units.}
    \label{fig: Evolution}
\end{figure}

In this subsection, we discuss the main results regarding stability. Fig.~\ref{fig: Evolution} displays the evolution of the models on one of the equilibrium sequences (sequence G). Here we show some representative models that were evolved for this sequence near the turning point. The maximum density normalised to its initial value against time is presented. The evolution was performed for 6\,ms. An inward radial velocity perturbation was applied (see Eq.~\ref{eq:perturbation}) on the stars at $t=0$. Notice that the stars with the higher energy densities collapse within the dynamical timescale ($\sim$1--2\,ms). The low energy density stars, on the other hand, oscillate about their equilibria but remain stable on the relevant timescale. All the stars that are stable fall on the low density side of the turning point thus obeying the turning point criterion. This feature can be observed in all the other sequences, where the higher energy density stars will collapse within $\sim$1--2\,ms and the lower density stars are stable and oscillate. All the stars on the left side of the turning point are not necessarily stable as pointed out by Weih~{\it et al}~\cite{Weih:2017mcw}. Since the actual onset of instability is marked by the neutral stability line which may lie to the left of the turning point~\cite{Takami:2011zc}, some stars on the left side of the turning point are unstable. Nevertheless, it can still be concluded that the instability is reached at or before the turning point making the turning point criterion a sufficient condition for instability. Our findings for all the sequences conform to this. The oscillations in the stable stars are due to the perturbation.  If the same star is evolved without perturbation, the oscillation amplitude is much smaller (although not exactly zero, due to truncation error in the initial data).

\begin{figure*}
\includegraphics[width=1\columnwidth]{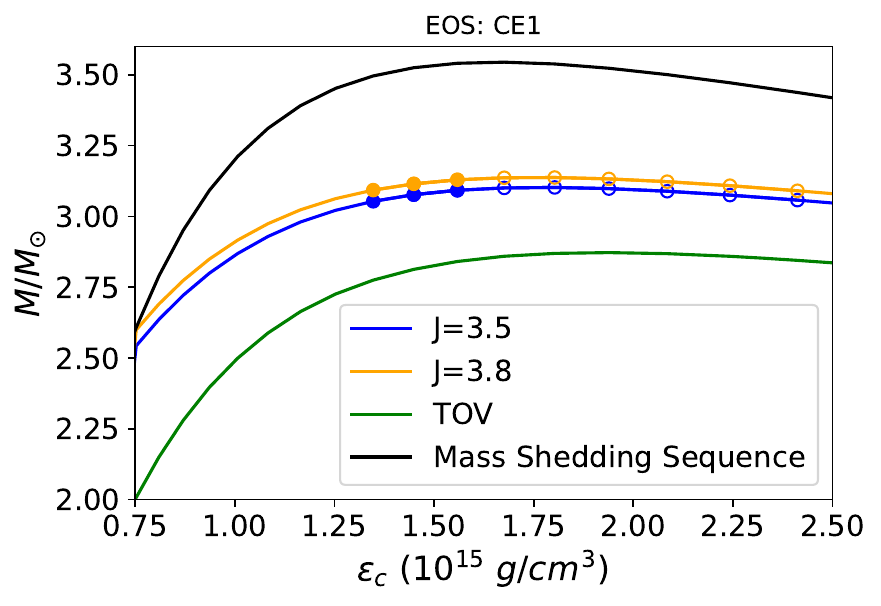}
\includegraphics[width=1\columnwidth]{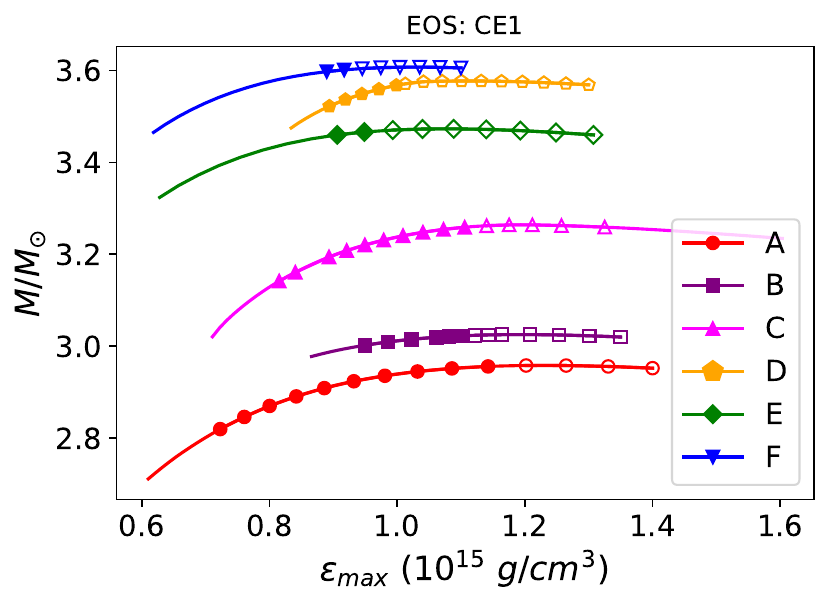} \\
\caption{Constant angular momentum sequence for CE1 entropy profile. {\bf Left panel}: Sequences with j-constant rotation law (Eq.~\ref{eq:j-const}). Apart from j-sequences the TOV and Mass-shedding sequences are shown; {\bf Right panel}: Sequences with Ury{\={u}} rotation law (Eq.~\ref{eq:uryu}).} 
\label{fig:seq1}
\end{figure*}

\begin{figure*}
\includegraphics[width=1\columnwidth]{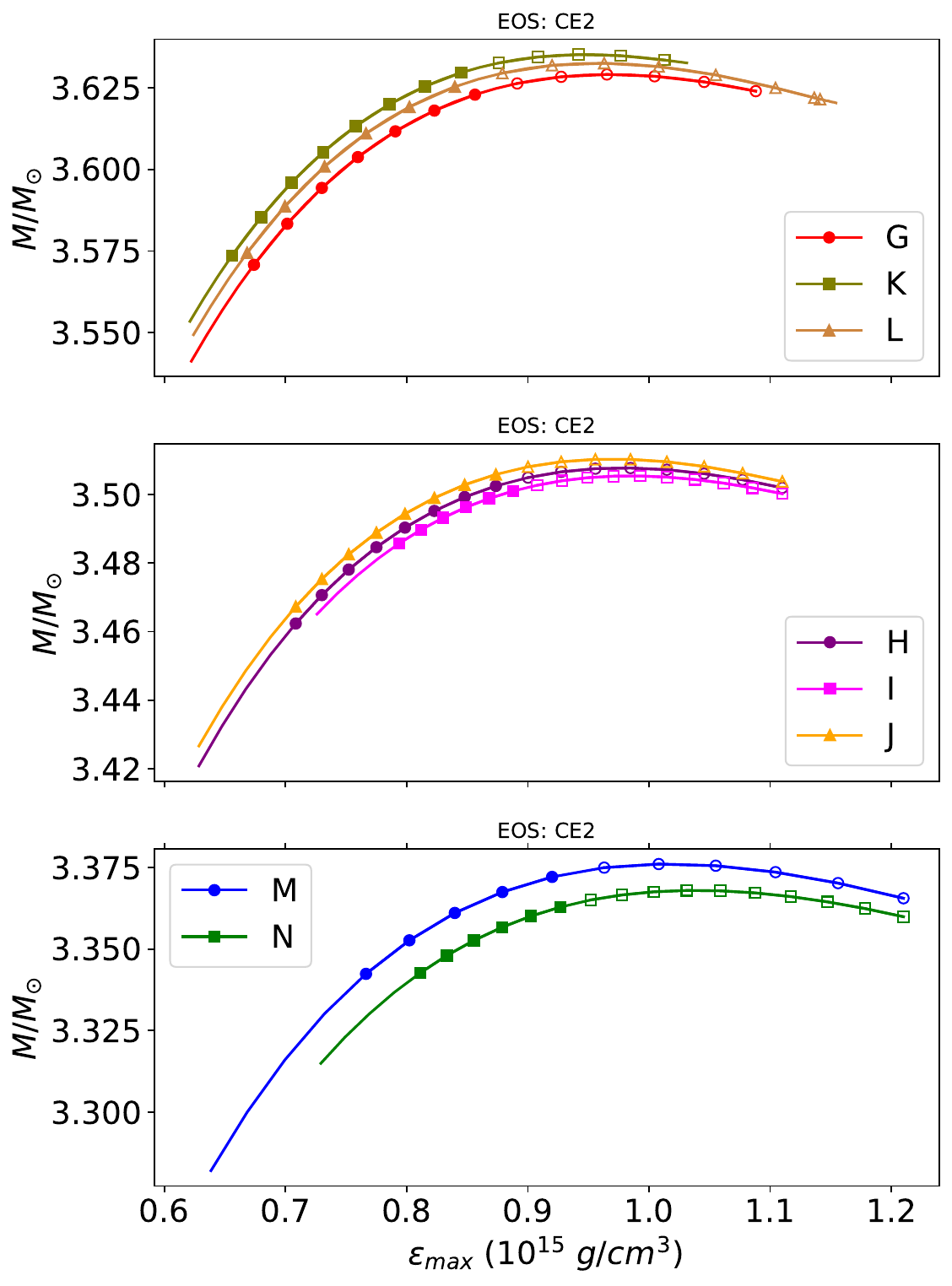}
\includegraphics[width=1\columnwidth]{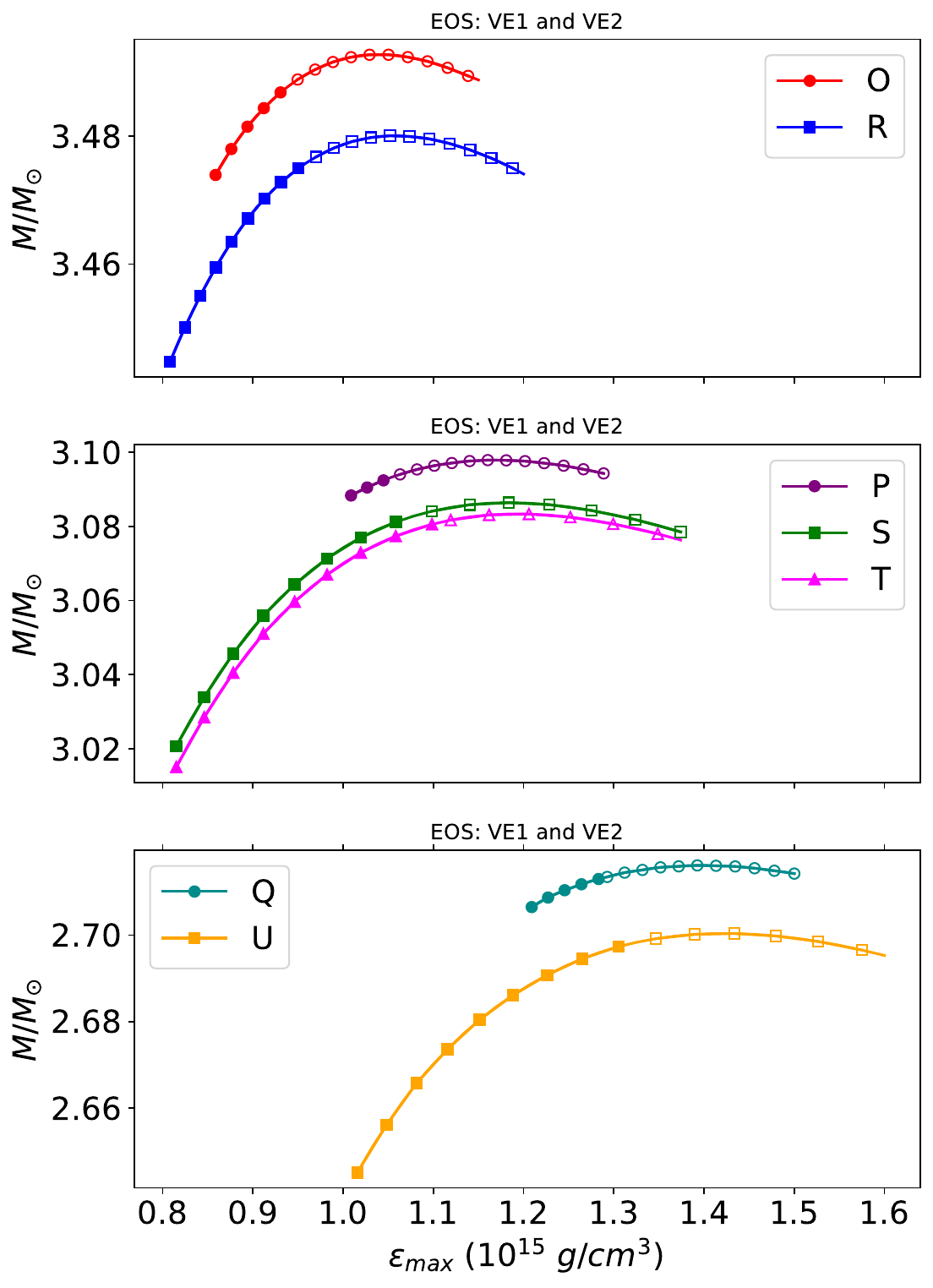} \\
\caption{Constant angular momentum sequences for Ury{\={u}} rotation law. {\bf Left panels}: Sequences with CE2 entropy profiles.; {\bf Right panels}: Sequences with variable entropy profiles (VE1 and VE2). The sequences are divided in three panels for the ease of visualization as they are clumped in three different regions.}
\label{fig:seq2}
\end{figure*}

What has been illustrated in figure~\ref{fig: Evolution} for one sequence can be succinctly presented in the sequence plots (see Fig.~\ref{fig:seq1} and~\ref{fig:seq2}), given that the focus of this study is the stability of these models. These figures show the gravitational mass vs the maximum energy density of constant angular momentum sequences except for the left panel of Fig.~\ref{fig:seq1}, where the x-axis represent the central energy density.

First, we reproduce the finding of Weih~{\it et al} for a monotonic rotation law for a small sample of cases. The left panel of Fig.~\ref{fig:seq1} shows the sequences for CE1 equation of state for j-constant rotation law. The top and the bottom curves represent the mass shedding and the TOV sequences respectively. The middle curves represent $J=3.5$ and $J=3.8$ sequences with a rotation parameter $A=0.2$. The filled and unfilled markers on all the sequence figures represent the stars that were evolved on that respective sequence. The filled shapes represent the stars that were stable after the evolution and the hollow ones represent the stars that collapsed into a black hole (the density blew up) within the dynamical timescale. We find that the sequences with j-constant law follow the turning point criterion. All the stable stars lie on the low density side of the turning point. This result is consistent with Weih~{\it et al}~\cite{Weih:2017mcw}'s findings.

In the right panel of Fig.~\ref{fig:seq1} the constant angular momentum sequences with the Ury{\={u}}~{\it et al} rotation law for the CE1 EOS cut are presented. The angular momenta of the sequences are 6, 6.5, 8.5, 10 and 11 (see Table~\ref{table:SequenceNames}). All the stable stars lie on the left of the turning point.  Furthermore, the transition from stable to unstable stars is close to the turning point, consistent with the approximate turning point conjecture of Kaplan~{\it et al}~\cite{Kaplan:2013wra}.

In Fig.~\ref{fig:seq2}, different panels show sequences with different 1D cuts of the 3D DD2 equation of states. We note that because these sequences have constant $s(\rho)$, they do not have constant total entropy or total entropy per baryon.  As $\rho_{\rm max}$ increases along a sequence, more of the low-entropy high density region is sampled, and the specific entropy averaged over the star decreases.  The three panels on the left show sequences of equation of state cut CE2. The three panels on the right show sequences with equation of state of variable entropy VE1 and VE2. The panels have been divided according to angular momentum ranges for better visualization. Each of the sequences are constructed with different angular momentum and different combinations of $A$ and $B$ (see Table~\ref{table:SequenceNames}). The $A$ and $B$ combinations are chosen to make neutron star merger-like profiles. It is reflected in Fig.~\ref{fig: OmegaProfileEvolution}, where the rotation profile of a star on sequence G is plotted. The rotation profiles, in particular the $\Omega_{\rm max}/\Omega_{\rm c}$ and the $\Omega_{\rm eq}/\Omega_{\rm c}$, of the stars are similar to that of a neutron star merger remnant. It should be noted here that the maximum mass of the TOV sequences for all the EOS cuts and Ury{\={u}} rotation law lies in the range 2.42 -- 2.44 $M_{\odot}$. Therefore, all the sequences here have maximum mass 20\% -- 50\% higher than the TOV mass indicating the stars on these sequences are hypermassive.  The stability of stars on these sequences also conform to the approximate turning point conjecture.

Constant angular momentum sequences with constant $A$ and $B$ are not the unique way to construct sequences with this 2-parameter rotation law.  This points to a certain ambiguity of application in the turning point criterion.  Suppose a star belongs to two sequences of the same $J$ but with different rotation profile parameter held fixed.  Could the star be on the stable branch of one and the unstable branch of the other?  We have therefore also evolved several consant $J$ sequences with constant $\Omega_{\rm max}/\Omega_{\rm c}$ and $\Omega_{\rm eq}/\Omega_c$.  The $M$ vs $\rho_{\rm max}$ plots for these sequences are very close to the sequences of constant $A$ and $B$ with which they share a common starting point, so naturally the approximate turning point method works equally well for them.  This is consistent with the assumption that $M$ does not depend on angular momentum distribution to first order.

\subsection{Dependence of $M_{\rm max}$ on $A$ and $B$}
\label{sec:MmaxVsAB}

\begin{figure}
    \centering
    \includegraphics[width=1\linewidth]{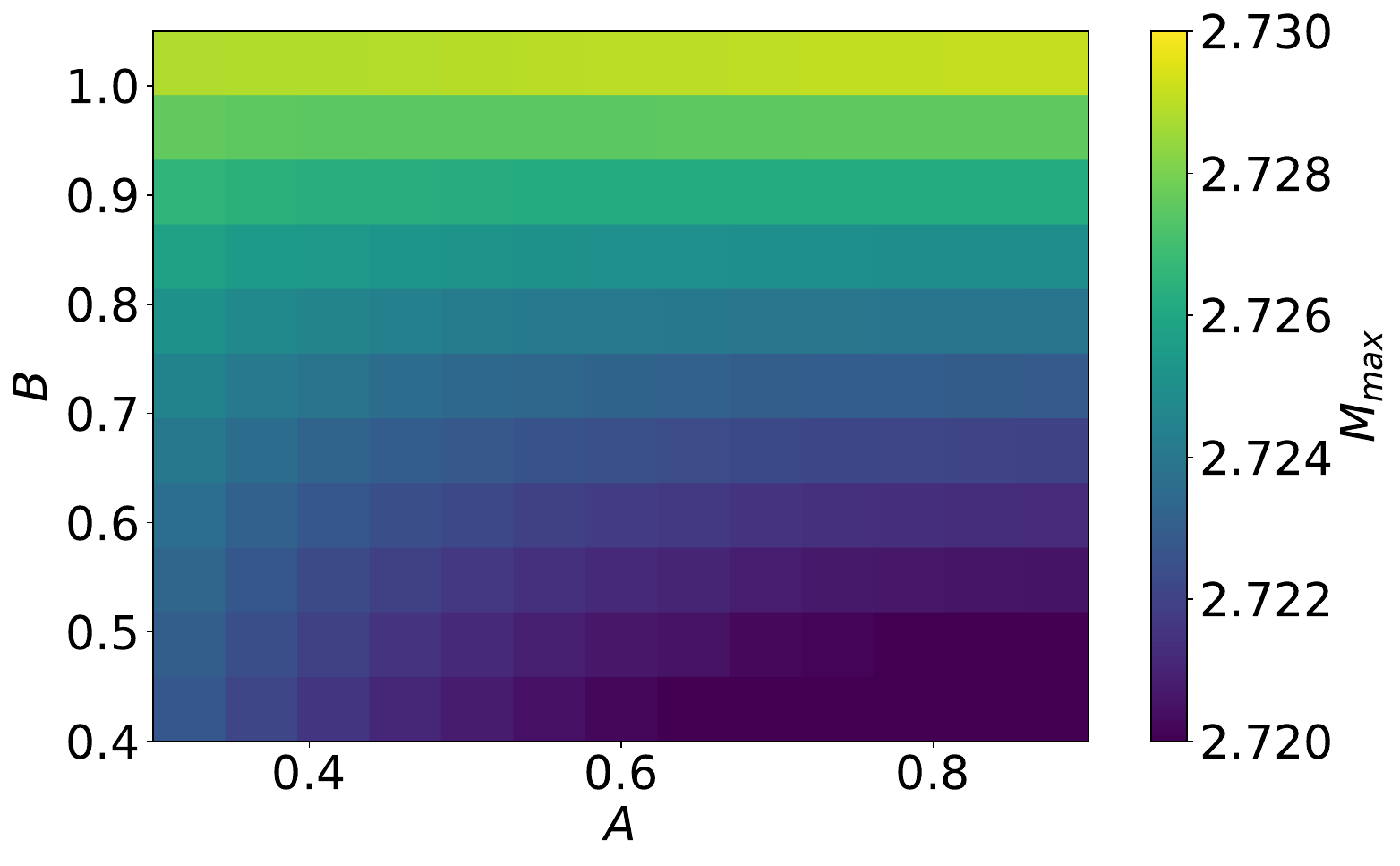}
    \caption{Maximum mass for $J=4$ sequences as a function of Ury{\=u} law rotation parameters $A$ and $B$.  }
    \label{fig: MmaxVsAB}
\end{figure}

It is already evident from the constant $J$ sequence plots that the maximum mass does not vary greatly with $A$ and $B$ parameters of the Ury{\={u}} law as compared with its variation along angular momentum or the EOS cuts.  This is consistent with the assumption of Kaplan~{\it et al} that $M_{\rm max}$ depends to first order only on the total angular momentum (and other global conserved quantities) and not its distribution.  We have systematically explored the dependence of the maximum mass for entropy profile CE2 on $A$ and $B$ for a fixed $J$  Fig.~\ref{fig: MmaxVsAB} illustrates the maximum mass of the sequences with $J = 4$ in the parameter space of $A$ and $B$ parameters of the Ury{\={u}} rotation law.  Note that this $J$ is lower than that of most of the hypermassive sequences studied above.  The plot shows not what is the greatest $M_{\rm max}$ attainable at any $J$, but which distribution of angular momentum gives the greatest enhancement of maximum mass for a given modest total angular momentum.  In this plot the parameter $A$ has been varied with an increment of 0.5, starting from 0.4 varying up to 0.9. The increment in $B$ is the same with a range 0.3 - 1.05. We find that the maximum mass is more sensitive to $B$ as compared to $A$ in this range.  Highest $M_{\rm max}$ for a given $J$ is obtained for high $B$.

\section{Summary and conclusions}
\label{sec:summary}
We have seen that the approximate turning point method successfully predicts stability for the range of entropy and non-monotonic rotation profiles studied, chosen to model post-merger equilibria more realistically than previously realized. No doubt it would be possible to devise rotation and entropy profiles that resemble post-merger remnants even more closely.  However, if the goal is to test turning point methods in extreme conditions under which they might fail, perhaps a better strategy would be to investigate {\it less} realistic profiles. For example, significantly wider exploration of entropy effects will confront the inconvenience that entropies high enough for a non-degenerate core will (assuming one insists on convectively stable $ds/d\rho<0$ profiles) have thermally supported envelopes with extended low density region.  This can, in fact, make the mass-shedding limit more severe~\cite{Kaplan:2013wra}.

In the process of carrying out this exploration of rotation law parameter space for hypermassive neutron stars with Ury{\={u}}~{\it et al} equilibria, we have shown that we can generalize the RotNS code to cover a fair approximation to the realistic range of post-merger remnant rotation and entropy profiles.  Investigations of the late-time ($\sim$ seconds) evolution of binary neutron star and black hole-neutron star mergers often resort to 2D axisymmetric simulations.  The initial data for these simulations has been either simple equilibria (e.g. constant entropy, j-constant rotation), which are artificial, or azimuthally averaged profiles of 3D mergers, and this averaging is a strong and sudden perturbation of the 3D system (even many that are ``roughly axisymmetric'') and can produce worrying transients. Axisymmetric equilibria with profiles extracted from merger simulations (e.g. a fit to $A$, $B$, and $s(\rho)$) might provide an attractive combination of the best of these two methods: arguably capturing as much realism from 3D merger profiles as 2D can accommodate while avoiding transients.

\begin{acknowledgments}

M.D. gratefully acknowledges support from the NSF through grant PHY-2110287 and through REU grant PHY-2050886. 
M.D. and F.F. gratefully acknowledge support from NASA through grant 80NSSC22K0719. F.F. gratefully acknowledge support from the Department of Energy, Office of Science, Office of Nuclear Physics, under contract number DE-AC02-05CH11231 and from the NSF through grant AST-2107932.  M.S. acknowledges funding from the Sherman Fairchild Foundation and by NSF Grants No. PHY-1708212, No. PHY-1708213, and No. OAC-1931266 at Caltech.  L.K. acknowledges funding from the Sherman Fairchild Foundation
and by NSF Grants No. PHY-1912081, No. PHY-2207342, and No. OAC-1931280 at Cornell.  Computations for this manuscript were performed on the Wheeler cluster at Caltech, supported by the Sherman Fairchild Foundation.

\end{acknowledgments}

%\nocite{*}

\bibliography{reference}% Produces the bibliography via BibTeX.


\end{document}